\RequirePackage{lineno}
\documentclass[a4paper]{jpconf}
\usepackage{graphicx}
\usepackage{amssymb,amsmath,array}
\usepackage{subfigure}
\usepackage[vcentering,dvips]{geometry}

\def\mean#1{\left<#1\right>}

\begin{document}

\title{Measurements of heavy quark production via single leptons at PHENIX}


\author{Donald Hornback (for the PHENIX Collaboration)}

\address{University of Tennessee - Department of Physics and Astronomy\\
Knoxville, TN 37996, USA}
\ead{dhornbac@utk.edu}

\begin{abstract}
The measurement of single leptons from the semi-leptonic decay of heavy-flavor hadrons has long been a means for studying heavy-quark production.  PHENIX has measured single muons in $pp$ collisions at forward rapidity and single electrons in both $pp$ and $AuAu$ collisions at mid-rapidity at $\sqrt{s_{NN}}$ = 200 GeV.  The most recent PHENIX single lepton results are presented in the context of state-of-the-art pQCD calculations.   An updated azimuthal anisotropy, $v_{2}(p_{T})$, measurement for heavy-flavor single electrons in $AuAu$ collisions is also presented. 
\end{abstract}

\section{Introduction}

With the ability to measure heavy-flavor production in two regions, muons at forward rapidity and electrons at mid-rapidity, PHENIX is positioned to study the rapidity evolution of heavy-flavor production cross sections.  In $pp$ collisions, heavy-flavor cross section measurements serve to test pQCD predictions and provide baseline comparisons for heavy-ion studies.  Measurement of the nuclear modification factor, $R_{AA}$, and azimuthal anisotropy, $v_{2}(p_{T}),$ for heavy-flavor production provides further information for the study the dense partonic matter produced in heavy-ion collisions at RHIC.
  

\section{PHENIX experiment and analyses}

PHENIX \cite{PHENIX:NIM} measures single electrons with two separate central arms, each with 90$^\circ$ azimuthal acceptance and pseudorapidity coverage of $\mid \!  \eta \!  \mid \, \le$ 0.35. Electrons are identified in the central arms using combined information from an electromagnetic calorimeter and a ring imaging \^Cerenkov detector.  Muons are measured using two separate muon spectrometers with full azimuthal coverage and covering the pseudorapidity range 1.2 $\le \; \mid \! \! \eta \!  \! \mid \; \le$ 2.4.  Muons are identified in the forward and backward directions through the use of Iarocci tubes interleaved between steel absorber plates.  Muon momentum determination is accomplished using cathode strip chambers inside a magnetic field. 

The measurement of single electron and muons from heavy-flavor is performed through the statistical subtraction of background sources, with the remaining yield attributed to open heavy flavor decays.  For single electrons, the primary backgrounds arise from $\pi^{\circ}$ and $\eta$ Dalitz decay and photon conversion.  A ``cocktail'' of electron spectra using directly measured background sources \cite{PHENIX:ppg063} is calculated using a Monte Carlo event generator of 
hadron decays.  Additionally, the ratio of conversion electrons is altered by the addition of a thin layer of conversion material between the vertex and detector for a portion of the run period.  The relative increase in conversions provides a direct measurement of this key background and serves as an independent cross check of the cocktail approach \cite{PPG:065}.

For single muons a background cocktail approach is also used.  The backgrounds in the PHENIX muon arms can be categorized in two ways: 1) muons which result from the weak decay of light hadrons before the first absorber material, 2) hadrons which penetrate the $\sim$1.5 m of steel absorber to reach the deepest layer in the muon arm.  Muons from heavy flavor meson decay originate $<$1 mm from the collision vertex, while yields of muons from hadron decay exhibit a linear collision vertex dependence before the first absorber material due to decay kinematics.  Identified hadron measurements do not exist for $\mean{y}$=1.65, so input spectra are constrained using measurements at mid-rapidity \cite{PHENIX:ppg063} and at forward rapidity $y \sim$ 2.2 and 3 \cite{brahms:y3}.  The hadron background cocktail is tuned to simultaneously match data for both: 1) measured unidentified stopped hadrons in the shallow muon identifier layers and, 2) the measured linear z-vertex distributions that indicate the fraction of hadronic decay in the sample of muon tracks \cite{PHENIX:ppg057}.  
  

\section{Discussion}

\begin{figure}[b]
\begin{minipage}{17pc}
\includegraphics[width=17pc]{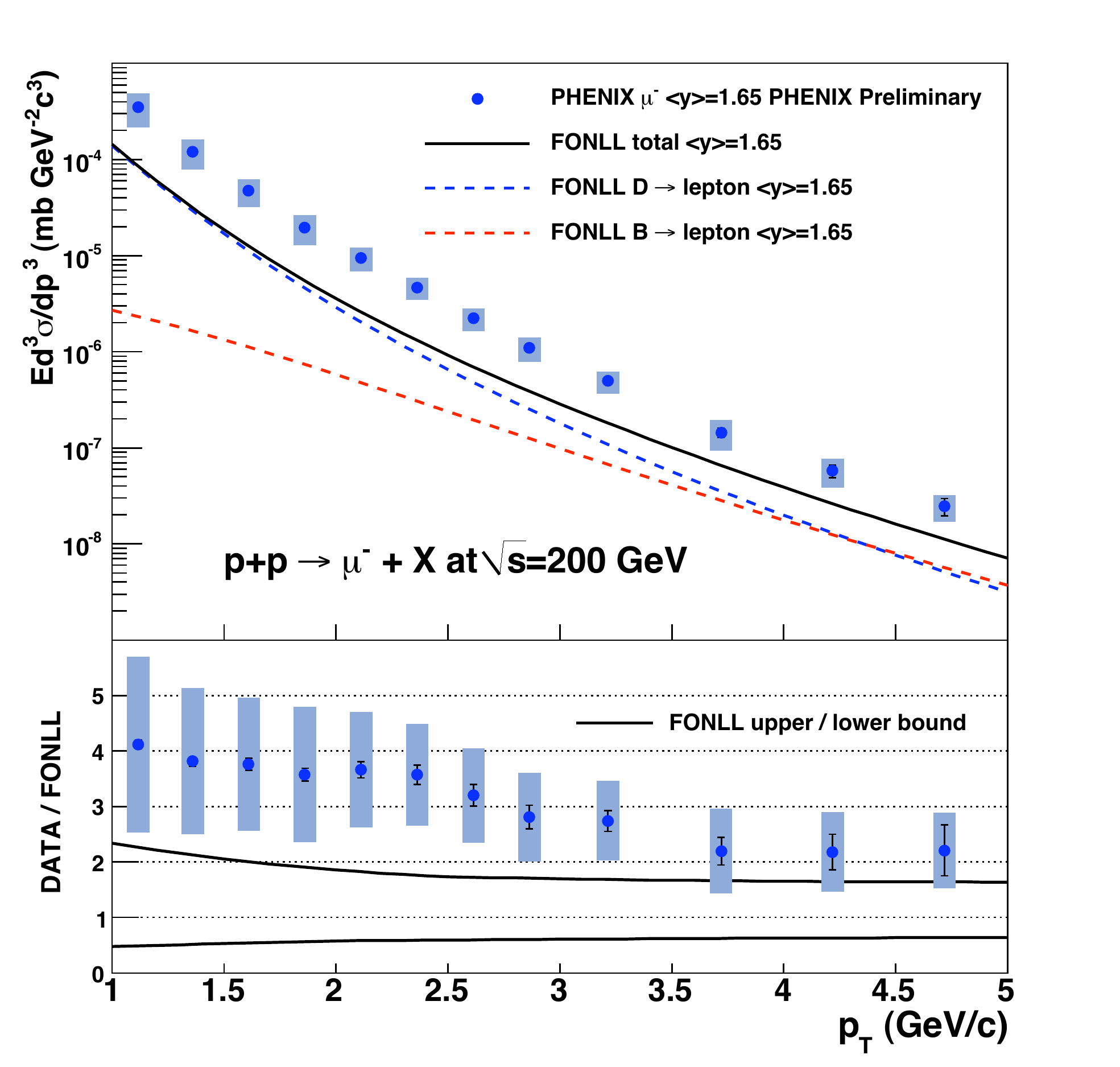}
\caption{\label{fig:muon_spectra}Upper plot: invariant differential cross sections of muons from heavy-flavor decay for $\mean{y}$=1.65.  The error bars (bands) represent the statistical (systematic) errors.  The curves are FONLL calculations.  Lower plot: ratio of data and FONLL calculation.  The upper (lower) curves represent the FONLL upper (lower) limits.}
\end{minipage}\hspace{2pc}%
\begin{minipage}{17pc}
\includegraphics[width=17pc]{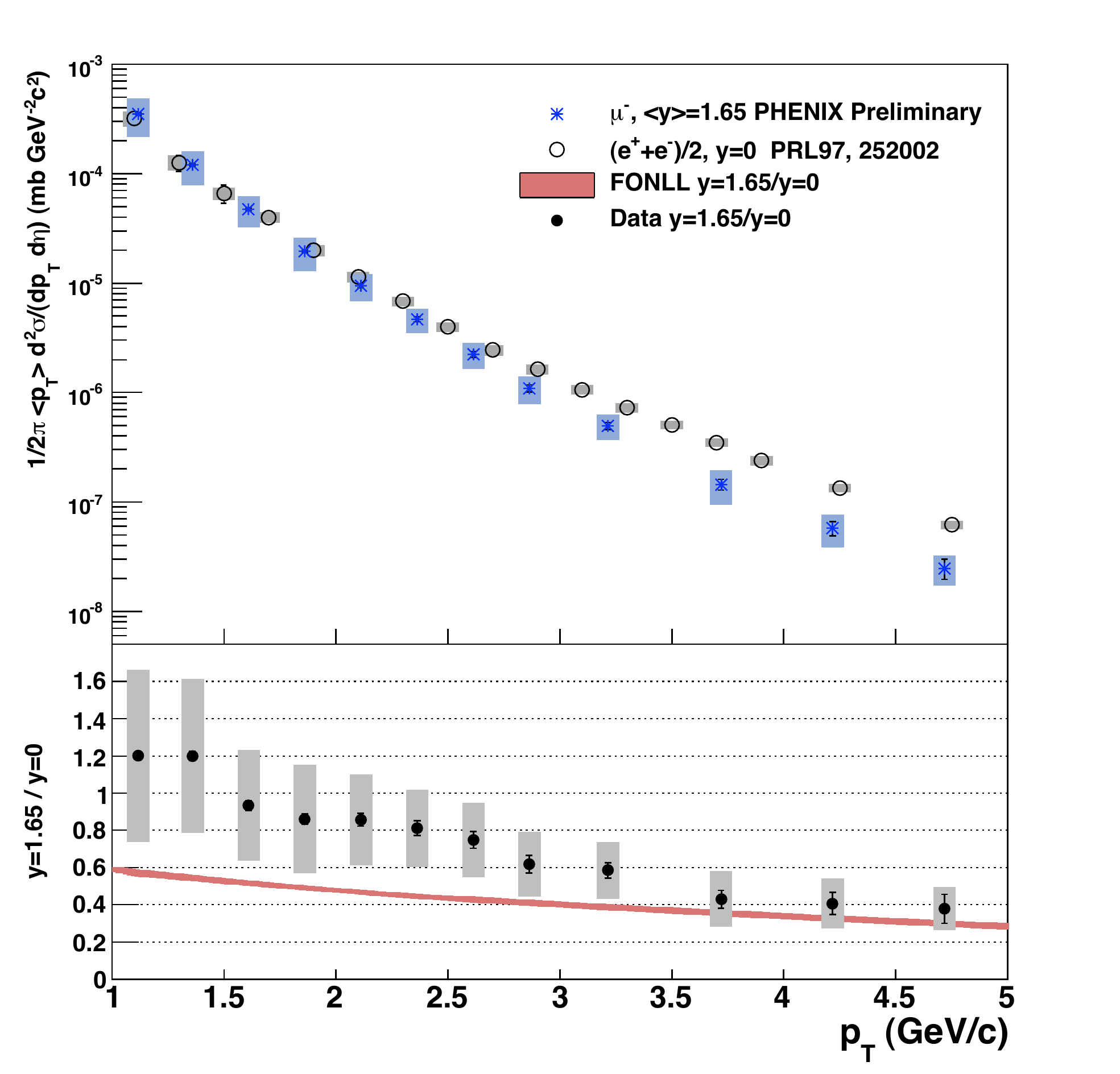}
\caption{\label{fig:muon_electron_ratio}Upper plot: invariant differential cross sections of muons from heavy-flavor decay at $\mean{y}$=1.65 and electrons at $\mean{y}$=0.  The expected softening of the spectrum at forward rapidity is apparent.  Lower plot: ratio of the forward muon spectrum to that of electron spectrum.  The solid red line is the y=1.65/y=0 ratio expected from the FONLL calculation. }
\end{minipage} 
\end{figure}

Within existing theoretical and experimental uncertainties bottom production is generally well described in $p\bar{p}$ collisions by pQCD calculations \cite{Matteo:2004}.  The situation in $pp$ collisions for charm is less sanguine on both the experimental and theoretical fronts.  The discrepancy between PHENIX and STAR single electron spectra is well known, with the STAR \cite{STAR:charm} measurement effectively a factor of two above that of PHENIX \cite{PPG:065} for all $p_{T}$.  Existing theoretical uncertainties on the total charm cross section, $\sigma_{c\bar{c}}$ span more than an order of magnitude \cite{Vogt:total}, and the uncertainty on the calculated $p_{T}$ spectra ranges from a factor of two at $p_{T} \approx$1.0 GeV/c to more than 50\% for $p_{T} \gtrsim$ 2.0 GeV/c \cite{FONLL:RHIC}.  The FONLL uncertainty band (bottom plot of Figure~\ref{fig:muon_spectra}) does not represent gaussian errors, rather the ``envelope'' of possible curves obtained through a systematic variation of the quark mass and the renormalization and factorization scales.  The band represents an approximately flat probability region with a high probability of containing the ``correct'' calculation \cite{FONLL:RHIC}.

PHENIX has previously published the first ever heavy-flavor single muon results in $\sqrt{s}$=200 GeV $pp$ collisions at  $\mean{y}$=1.65 \cite{PHENIX:ppg057}.  New PHENIX preliminary single muon spectrum is measured at $\mean{y}$=1.65 over the range 1.0 $\le p_{T} \le$ 5.0 GeV/c, extending the previous $p_{T}$ reach with improved statistics and reduced systematic uncertainties.  Single muons are measured in both forward and backward directions, with the two spectra being combined to form the spectrum shown in Figure~\ref{fig:muon_spectra}, which is compared to a FONLL calculation over the same rapidity \cite{FONLL:private}.  

\begin{figure}[bt]
\begin{minipage}{17pc}
\includegraphics[width=17pc]{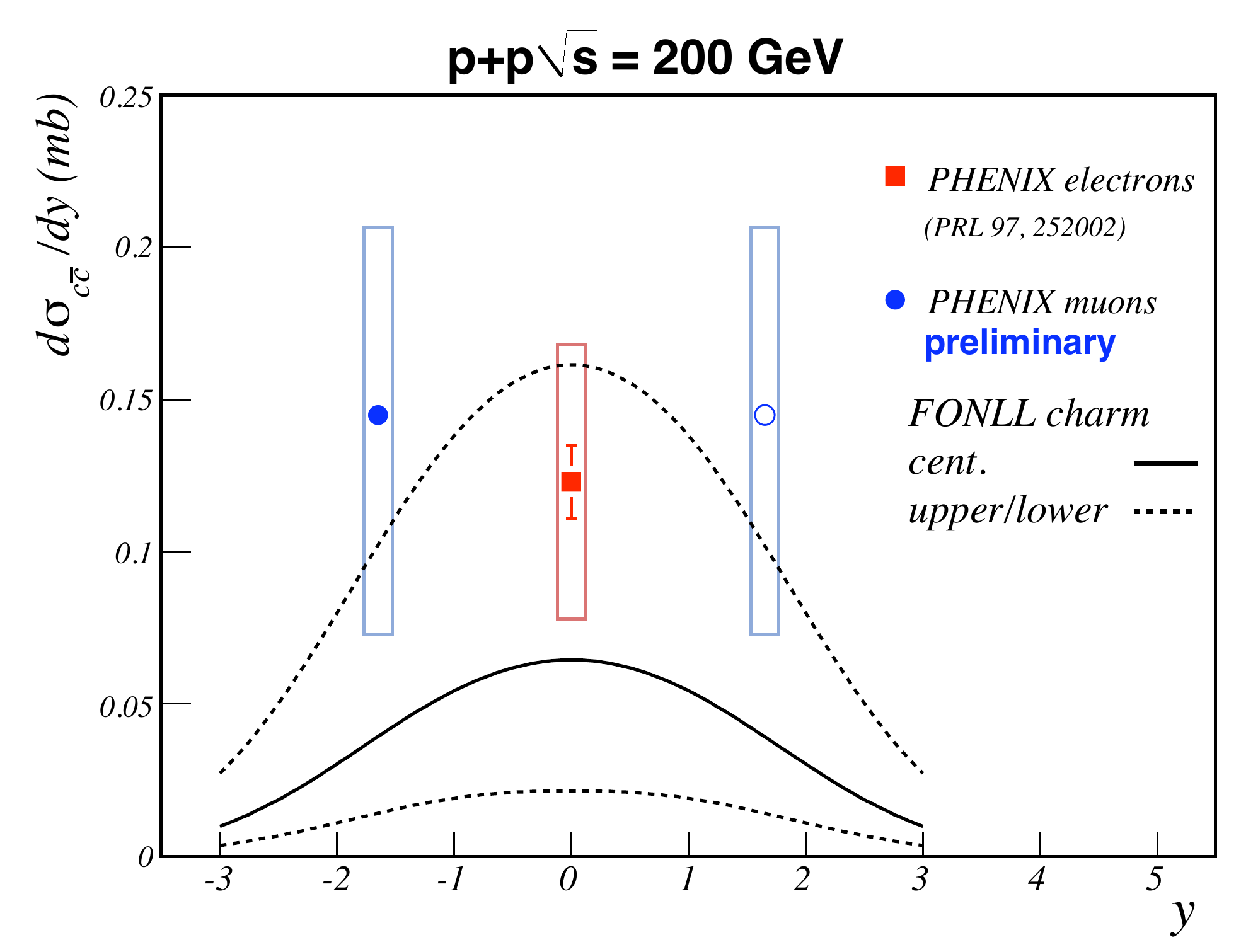}
\caption{\label{fig:integrated}Comparisons of measured charm cross sections, $d\sigma_{c\bar{c}} / dy$ at $\mean{y}$=0 and $\pm$1.65 compared to FONLL.  Within uncertainties, the PHENIX y=1.65 point (circle) is consistent with the PHENIX y=0 point (square) and the upper bound of FONLL.}
\end{minipage}\hspace{2pc}%
\begin{minipage}{17pc}
\includegraphics[width=17pc]{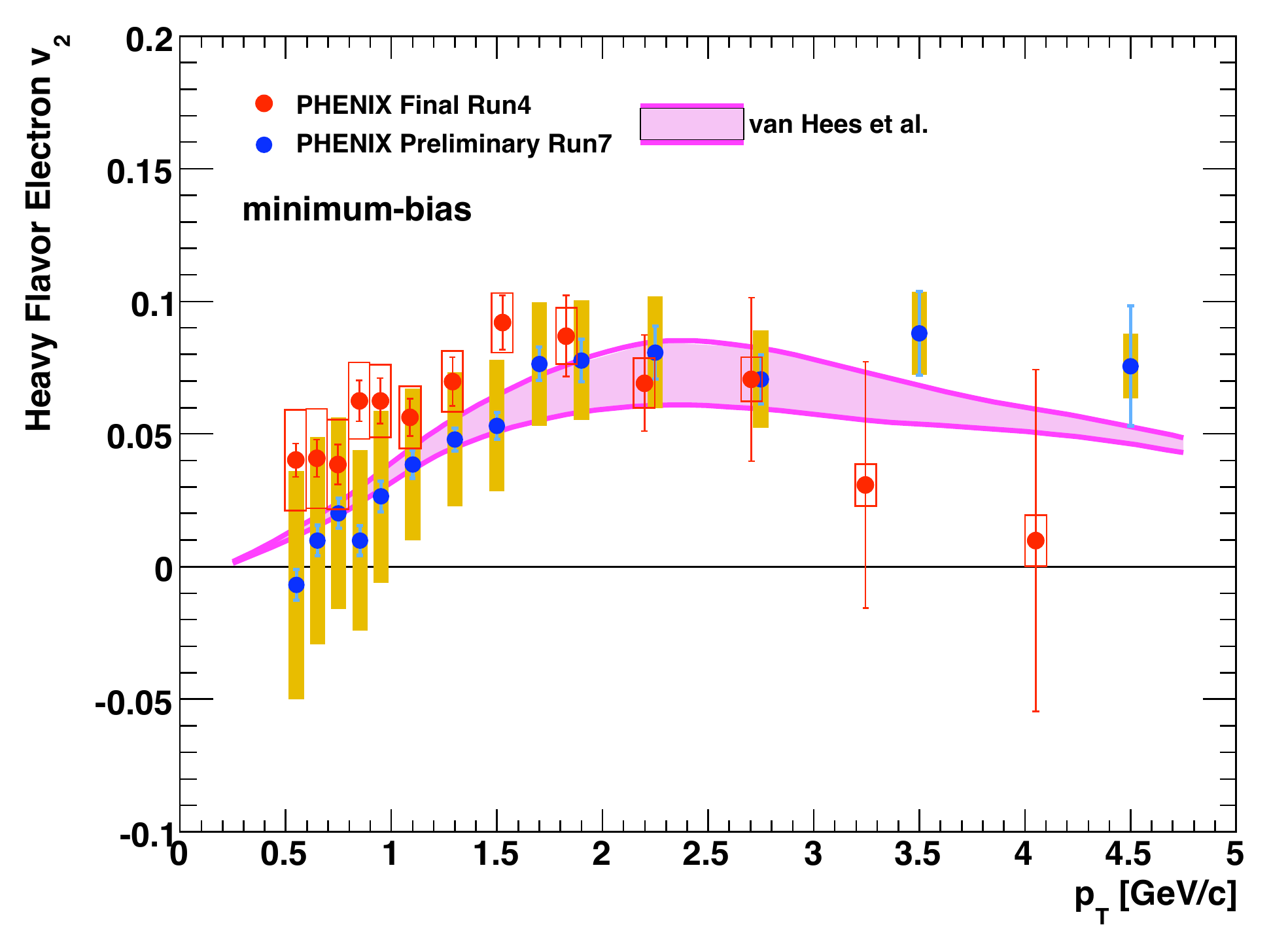}
\caption{\label{fig:v2}PHENIX $v_{2}$ of heavy-flavor electrons in minimum bias collisions from Run 4 \cite{PPG:066} and preliminary Run 7 compared to the transport calculation by van Hees et. al. \cite{vanHees}. }
\end{minipage} 
\end{figure}

Relative to the $\mean{y}$=1.65 FONLL calculation, the single muon spectrum is consistent within uncertainties for $p_{T}>$ 3.5 GeV/c where the experimental signal/background (S/B) ratio is highest and where the FONLL calculation is most reliable \cite{Vogt:total}.  For $p_{T}<$ 3.5 GeV/c the single muon data sits above the FONLL central curve by a factor of $\sim$3.5.  The sizable systematic uncertainties $\sim$30-40\% arise primarily from the S/B of $\sim$0.3, where a $\sim$10\% underestimate of the total background would result in a $\sim$30\% increase in signal.  The previously reported PHENIX single electron spectrum \cite{PPG:065} over 0.3 $<$ $p_{T}$ $<$ 9.0 GeV/$c$ lies along the upper theoretical uncertainty band and is therefore consistent with the FONLL calculation.  This electron spectrum has a constant-to-fit of 1.72$\pm$0.02(stat)$\pm$0.19(sys) relative to the respective $\mean{y}$=0 FONLL calculation  \cite{PPG:065}.  The upper plot of Figure~\ref{fig:muon_electron_ratio} shows that the $\mean{y}$=1.65 spectrum exhibits a softer shape when compared to $\mean{y}$=0, as expected, though with decreasing $p_{T}$ the muon spectrum trends above the y=1.65/y=0 ratio expected from FONLL (lower plot).  While the single electron spectrum is approximately constant in $p_{T}$ relative to the FONLL calculation, the single muon data trends downward with increasing $p_{T}$.  This effect may possibly be explained through some combination of a systematic effect in the background estimation and a difference in the rapidity fall off expected by FONLL and present in the data (Figure~\ref{fig:integrated}).

An integrated charm cross section at a particular rapidity can be obtained by extrapolation of the $p_{T}$ spectra to $p_{T}$=0 GeV/c by some means such as FONLL.  At $\mean{y}$=0, PHENIX measures a $d\sigma_{c\bar{c}}/dy \mid_{y=0}$ = 0.123$\pm$0.012 (stat.)$\pm$0.45 (sys.) mb \cite{PPG:065}, and STAR measures $d\sigma_{c\bar{c}}/dy \mid_{y=0}$ = 0.30$\pm$0.04$\pm$0.09 mb \cite{STAR:Ds}.  Performing the same exercise on the single muon spectra yields a charm cross section of $d\sigma_{c\bar{c}}/dy \mid_{y=1.65}$ = 0.145$\pm$0.002$^{+ 0.062}_{- 0.072}$ mb.  This result is consistent with the previously reported charm cross section from single muons at $d\sigma_{c\bar{c}}/dy \mid_{y=1.65}$ = 0.243$\pm$0.013$^{+ 0.116}_{- 0.136}$ mb.  The reduction of the central point from 0.243 to 0.148 mb is due primarily to a change from PYTHIA to FONLL in the extrapolation of the spectra for $p_{T}<$1.0 GeV/c.  The total charm cross section determined from the PHENIX $\mean{y}$=0 charm measurement is determined to be $\sigma_{c\bar{c}}$ = 567 $\pm$ 57$^{stat}$ $\pm$ 224$^{sys}$ $\mu b$.  The uncertainties on the single muon point are too large at this time to constrain the rapidity shape of the charm cross section.  

Observations by both PHENIX \cite{PPG:066} and STAR \cite{STAR:charm} of the $R_{AA}$ are in agreement in Au+Au collisions showing the suppression of heavy quarks at high $p_{T}$.  This phenomena is interpreted as arising from larger than expected partonic energy loss of heavy quarks in the medium.  PHENIX has reported an updated $v_{2}$ measurement from heavy-flavor single electrons  in $AuAu$ collisions.  The new measurement benefits from the new reaction plane detector which provides an improved reaction plane resolution by a factor of 1.8.  However the new measurement also suffers from larger systematic errors than the previous result due to running with a reduced magnetic field strength and without a helium bag which increased the photonic conversion in air.  Figure~\ref{fig:v2} shows the new preliminary result compared to the previous measurement and the transport calculation by van Hees et. al. \cite{vanHees}.  The updated results are consistent within errors with the previous measurement, while the clear non-zero $v_{2}$ value is observed at high $p_{T}$ where dominant contribution from bottom quarks is expected to reduce both the observed suppression in $R_{AA}$ and non-zero $v_{2}$.




%

\section*{References}


\bibliographystyle{h-physrev3.bst}
\bibliography{hornback_donald}



\end{document}